\documentclass[lettersize,journal]{IEEEtran}
\usepackage[T1]{fontenc}
\usepackage[utf8]{inputenc}
\usepackage{amsmath,amsfonts}
\usepackage{algorithmic}
\usepackage{array}
\usepackage[caption=false,font=normalsize,labelfont=sf,textfont=sf]{subfig}
\usepackage{textcomp}
\usepackage{stfloats}
\usepackage{url}
\usepackage{verbatim}
\usepackage{graphicx}
\def\BibTeX{{\rm B\kern-.05em{\sc i\kern-.025em b}\kern-.08em
    T\kern-.1667em\lower.7ex\hbox{E}\kern-.125emX}}
\usepackage{balance}
\usepackage[table]{xcolor}
\usepackage{booktabs}
\definecolor{orangeheader}{RGB}{204, 102, 51}
\definecolor{orangelight}{RGB}{255, 228, 196}
\definecolor{blueheader}{RGB}{51, 102, 153}
\definecolor{bluelight}{RGB}{220, 230, 255}

\definecolor{greenheader}{RGB}{46, 125, 50}
\definecolor{greenlight}{RGB}{200, 230, 201}
\definecolor{greenheader}{RGB}{34, 139, 34}
\definecolor{greenlight}{RGB}{220, 255, 220}
\definecolor{bluelight}{RGB}{230, 242, 255}
\definecolor{greenlight}{RGB}{232, 245, 233}
\definecolor{bluelight}{RGB}{227, 242, 253}

\begin{document}

\title{HAPS‑ISAC for 6G: Architecture, Design Trade‑offs, and a Practical Roadmap}
\author{Parisa Kanani, Mohammad Javad Omidi, Mahmoud Modarres-Hashemi,\\
	and Halim Yanikomeroglu,~\IEEEmembership{Fellow,~IEEE}

	\thanks{P. Kanani, M. J. Omidi, and M. Modarres-Hashemi are with the 
		Department of Electrical and Computer Engineering,
		Isfahan University of Technology, Isfahan 84156-83111, Iran (emails: p.kanani@ec.iut.ac.ir; omidi@iut.ac.ir; modarres@iut.ac.ir).}%
	\thanks{M. J. Omidi is also with the Department of Electronics and Communication Engineering, Kuwait College of Science and Technology, Doha 35003, Kuwait.}%
	\thanks{H. Yanikomeroglu is with Non-Terrestrial Networks (NTN) Lab, Department of Systems and Computer Engineering, Carleton University, Ottawa, ON K1S 5B6, Canada (email: halim@sce.carleton.ca).}%
}


\maketitle

\begin{abstract}
To meet the ambitious goals of next-generation 6G networks, including ultra-high data rates and ubiquitous coverage, we propose a novel high-altitude platform station (HAPS)-based integrated sensing and communication (ISAC) architecture. Operating in the stratosphere, the HAPS functions as both a powerful communication hub and an advanced environmental sensor. Combined with a fleet of cooperative uncrewed aerial vehicles (UAVs), this dual-purpose system forms a scalable and intelligent 3D network. Simulation results indicate that this approach significantly boosts network performance, improves sensing accuracy, and ensures a fairer service distribution across users, outperforming conventional UAV-only baselines. We conclude by outlining the prospective applications and a deployment roadmap for this technology for smart cities and other large-scale environments.
	
	

\end{abstract}


\section{Introduction}
\IEEEPARstart{T}{he} relentless surge in demand from transformative applications is exposing fundamental limitations of current fifth-generation (5G) networks. These applications—such as those in vehicular systems (e.g., autonomous driving, vehicle-to-everything (V2X) communications, intelligent transportation networks, and real-time traffic optimization)—demand extraordinary multi-terabit-per-second throughput, sub-millisecond latency, near-perfect reliability, and ubiquitous coverage. This unprecedented need serves as the primary catalyst driving the global research community toward sixth-generation (6G) wireless systems. More than an incremental upgrade, 6G represents a reimagined infrastructure featuring integrated terrestrial–non-terrestrial networks, sub-THz spectrum utilization, cell-free massive multiple-input multiple-output (MIMO), reconfigurable intelligent surfaces, and artificial intelligence (AI)-native resource management.
These technologies, taken together, are poised to transform connectivity across sectors—for example, by enabling next-generation intelligent transportation systems (ITS) \cite{veic2,veic1}.

%


In response, non-terrestrial networks (NTNs)—comprising spaceborne platforms (e.g., satellites)
and airborne platforms (e.g., high-altitude platform stations (HAPS) and uncrewed aerial vehicles
(UAVs))—have emerged to complement terrestrial infrastructures. These systems are pivotal for
extending coverage to underserved areas, bridging the digital divide, and ensuring disaster-resilient
communications, thereby realizing the pervasive and reliable 6G vision \cite{222,1}. Among NTN components,
HAPS stands out as a uniquely promising technology. Positioned quasi-stationarily in the
stratosphere (20–50 km altitude), HAPS offers distinct advantages over both terrestrial and satellite
networks. Compared to satellites, HAPS features significantly lower signal propagation delays
and substantially reduced operational and
deployment costs. Unlike UAVs, which are constrained by battery life and operational stability,
HAPS can be continuously powered, making them ideal for long-duration, large-scale missions.
Their extensive line-of-sight (LoS) visibility further enables wide-area coverage with stable and
reliable signals \cite{222,veic1,777}.

At the same time, integrated sensing and communication (ISAC) has emerged as a promising paradigm for 6G. ISAC integrates critical functionalities such as radar sensing and localization with wireless communication into a unified system. This integration significantly enhances efficiency and reduces costs by enabling both functionalities to share the same hardware and spectral resources
\cite{man,9,777}.

Terrestrial ISAC deployments face significant constraints. Physical obstacles often block direct line-of-sight (LoS) links, thereby impairing sensing tasks such as target detection and localization. Moreover, achieving precise long-range sensing requires substantial power consumption, which can diminish the communication performance of terrestrial base stations (BSs)  \cite{9,6}.

 While satellite- and UAV-based ISAC schemes have been explored, existing
research remains predominantly ground-centric, leaving the unique potential of HAPS to deliver
stable, high-precision ISAC across diverse environments largely untapped. In stark contrast,
integrating ISAC with HAPS offers distinct advantages: it can extend coverage, improve resource
allocation, and optimize beamforming for both sensing and communication. HAPS provides wider
coverage than ground BSs and significantly lower latency than low Earth orbit (LEO) satellites,
making it well-suited for densely populated urban areas. Moreover, unlike UAVs, HAPS ensures
sustained, reliable service over vast regions \cite{CPU_H,24,1,222}.

Motivated by this critical gap, we propose a novel HAPS-ISAC framework designed to
simultaneously achieve wide-area coverage, ultra-low latency, and fair resource allocation, thereby
significantly enhancing spectral efficiency, signal quality, and end-to-end latency in future 6G
networks. Integrating ISAC within HAPS architectures presents substantial technical challenges
due to the inherently multidimensional, nonlinear, and non-convex nature of the underlying
optimization problems, often categorized as NP-hard. These complexities stem from the necessity
to jointly manage communication quality, sensing accuracy, interference mitigation, and resource
allocation in dynamic, wide-area environments. To effectively address these challenges, we
leverage genetic algorithm (GA)-based optimization techniques, renowned for their global search
capability, adaptability, and robustness in avoiding local optima  \cite{ga_33,gahaps2}. Complementing these
algorithmic approaches, advanced MIMO and multiple-input single-output (MISO) beamforming techniques play a pivotal role in
overcoming the physical-layer constraints of HAPS-ISAC. By enabling directional transmission
and spatial selectivity, beamforming enhances link quality, improves sensing resolution, and minimizes interference, facilitating simultaneous service to multiple users and targets. These
strategies can dynamically adapt to changing network conditions and user distributions, ensuring
both reliable communication and accurate target tracking \cite{777,MISO2}. Collectively, these methodologies form
a coherent and efficient response to the critical challenges in deploying HAPS-ISAC systems at
scale.


This article introduces a novel HAPS-ISAC framework designed to meet the dual demands for enhanced communication and high-precision sensing in 6G networks. As depicted in Fig. 1, our proposed architecture seamlessly integrates multiple layers of connectivity, including terrestrial base stations, satellites, and HAPS, to create a robust and resilient network. This multi-tiered system ensures reliable communication in diverse environments, ranging from dense urban centers to remote and agricultural areas, while facilitating advanced sensing capabilities for various applications, such as autonomous vehicles.
Services are provisioned to users and targets either directly through HAPS or indirectly
via UAVs. 

The remainder of this paper is structured as follows: Section II details the HAPS-ISAC
architecture and its performance advantages for 6G, focusing on integrated sensing and
communication functions. Section III discusses HAPS’ dual role of ISAC as a super macro base station (SMBS) and
central processing unit (CPU). Section IV addresses HAPS-ISAC’s optimization complexities
and proposes solutions. Section V outlines HAPS-ISAC’s potential applications. Section VI
presents two case studies: HAPS as a macro base station and HAPS as a CPU in a HAPS-UAV
ISAC system. Finally, Section VII provides the discussion, followed by the conclusion in Section VIII.

\section{HAPS-ISAC: Architecture and Performance Advantages for 6G Networks}
\label{hapsisac33}
	\begin{figure}[t]
	\centering
\includegraphics[width=0.5\textwidth]{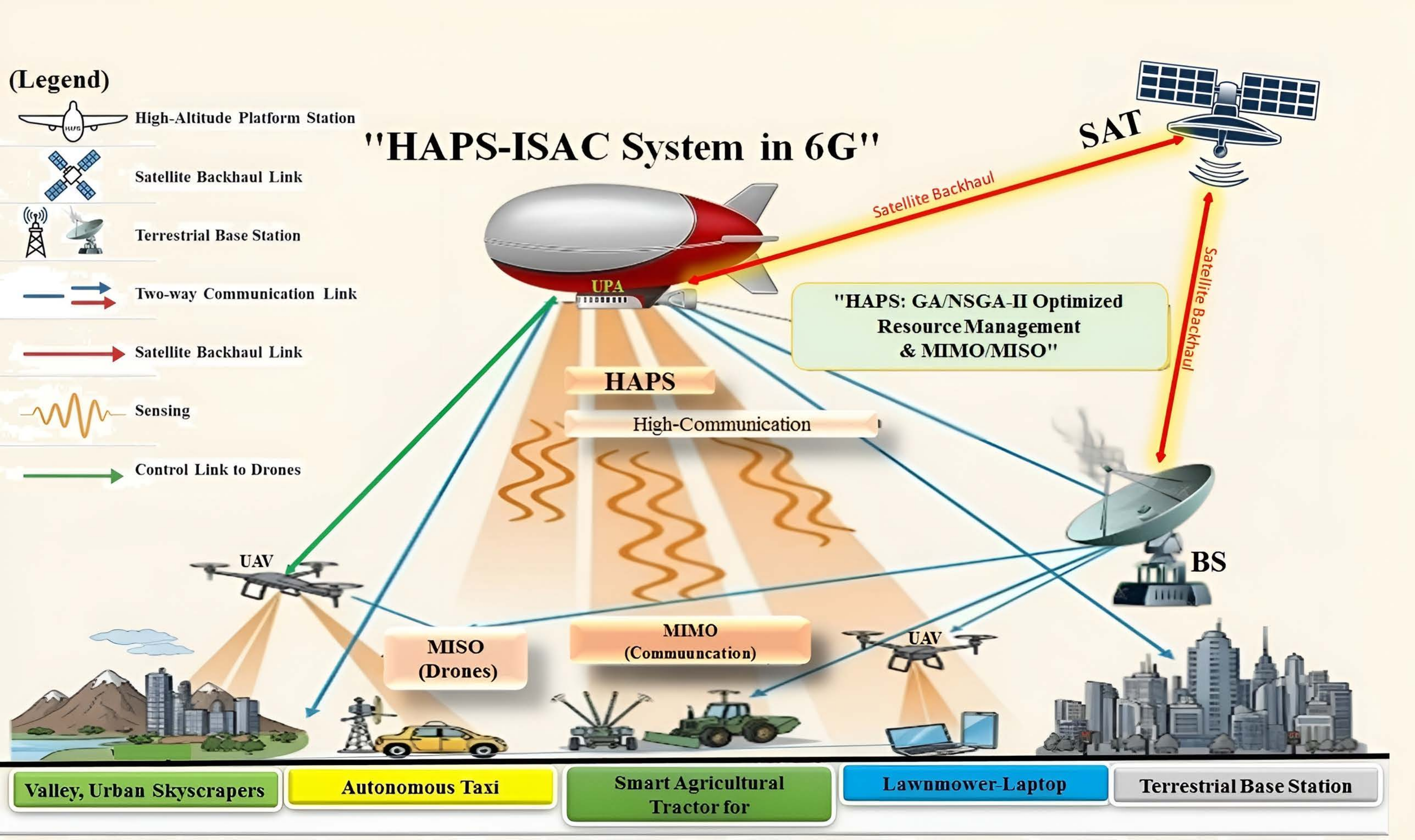}
	\caption{ HAPS-ISAC system architecture in 6G networks, highlighting HAPS as a core platform for integrated communication and sensing, crucial for enhancing network performance.}
	\label{systemba}
\end{figure}
The integration of HAPS with ISAC offers significant performance
improvements for 6G networks, notably addressing terrestrial infrastructure limitations through
wide coverage. HAPS provides superior stability and supports long-term operations over
extensive areas, unlike UAVs. This HAPS-ISAC synergy also significantly reduces latency and
enhances sensing resolution compared to satellite-based systems, while simplifying maintenance
and improving overall system efficiency. The unified communication and sensing capabilities of
HAPS-ISAC elevate target detection accuracy and service quality, establishing it as a
foundational model for future advancements in critical applications such as environmental
monitoring, disaster management, and emergency communications.
To illustrate these capabilities, Fig.~\ref{systemba} depicts the proposed multi-layered HAPS-ISAC architecture that integrates space, air, and terrestrial segments to provide ubiquitous connectivity and sensing. At the apex, geostationary or LEO satellites offer high-capacity backhaul, while the HAPS platform operates as a central aerial hub. This hub concurrently serves communication users (CUs)—leveraging MISO for energy-efficient single-antenna devices and MIMO for high-throughput multi-antenna systems—and performs sensing operations. The sensing framework is twofold: a \textit{device-free} mode for passive targets and a \textit{device-based} mode for cooperative targets (e.g., autonomous vehicles) that enable inter-entity communication. Seamless integration with terrestrial base stations ensures service continuity and load balancing. To orchestrate this heterogeneous environment, advanced optimization heuristics like the GA and non-dominated sorting genetic algorithm II (NSGA-II) are employed for dynamic resource management.

 \section{The dual role of HAPS  ISAC systems}
\label{basemacro}

This section explains the two primary roles of HAPS within the ISAC system. First, it operates as a large, high-powered cell tower—known as a super macro base station (SMBS)—that provides wide-area wireless coverage \cite{CPU_H}. Second, it functions as the network's central brain—the central processing unit (CPU)—coordinating all communication and sensing tasks. These dual roles are essential for enhancing the performance and integration of sensing and communication.

	\subsection{Deployment of HAPS as Super Macro Base Stations}
Utilizing HAPS as SMBS significantly enhances ISAC systems by optimizing energy consumption and improving overall network performance. The HAPS-based monostatic ISAC architecture integrates radar sensing and communication functions, enabling low-latency, high-bandwidth communication in remote rural areas and dense urban centers. Its stratospheric altitude ensures reliable operation, largely unaffected by atmospheric conditions, making it an effective alternative to terrestrial networks—particularly in emergency scenarios or areas with limited infrastructure.

HAPS supports diverse applications as macro base stations within ISAC frameworks. For instance, a monostatic HAPS-ISAC system can concurrently provide downlink communication for single-antenna users while performing radar-based sensing of potential targets. By employing Uniform Planar Array (UPA) steering vectors for beamforming, HAPS enhances both sensing resolution and communication efficiency, while adhering to signal-to-interference-plus-noise ratio (SINR) and power constraints. This beamforming capability enables precise multi-directional signal steering, thereby optimizing system performance under complex and dynamic conditions.
	\subsection{Deployment of HAPS as Central Processing Unit}
	HAPS is an ideal option for serving as a CPU due to its ability to maintain strong and reliable line-of-sight (LoS) links with remote platforms, minimizing blockage and shadowing in backhaul connections.
	As depicted in Fig. 1, the HAPS platform serves as a central processing and control hub across multiple tiers. It not only processes sensing data relayed from UAVs and manages communication for ground CUs (e.g., autonomous taxis and smart agricultural tractors) but also handles high-capacity satellite backhaul and integrates with terrestrial base stations. This centralized role is fundamental to network resilience; it enables dynamic traffic rerouting via satellite during disaster scenarios and allows for computational offloading from energy-constrained UAVs. Such offloading extends their operational endurance and significantly enhances the overall network's sustainability and efficiency.

	 This makes HAPS especially useful for backhauling aerial access points (APs) deployed in remote or underserved regions where terrestrial infrastructure may be insufficient or compromised.
	For instance, consider a scenario where an ISAC model is enhanced by a MIMO-capable HAPS-UAV system.
	In this case, HAPS serves as an aerial CPU responsible for backhauling and processing signals. The UAVs communicate directly with ground users and targets, while HAPS connects indirectly with these entities through the UAVs. To ensure accurate sensing of potential targets and improve signal reception for multiple CUs, beamforming techniques can be employed, utilizing steering vectors from a UPA at the UAVs to align incoming signals. By offloading computational tasks to HAPS, the UAVs can conserve energy, thereby enhancing their efficiency and the overall performance of the network.
	\section{Enhancing HAPS-ISAC Performance Through Optimization Algorithms}
	\label{alg}
A key challenge in ISAC systems is the need to jointly optimize both sensing and communication functionalities. This involves solving what is known as a multi-objective problem: a complex task that seeks to balance competing goals, such as maximizing user throughput (i.e., data speeds) while ensuring robust sensing performance. This optimization is particularly complex in HAPS-based systems due to their large multi-antenna arrays and the need for sophisticated beamforming techniques. Therefore, developing efficient optimization strategies is essential to manage these trade-offs effectively.

	\subsection{Role and Effectiveness of Genetic Algorithms in HAPS-ISAC Systems}
The optimization problems in HAPS–ISAC are typically non-convex and often NP-hard, so conventional methods struggle to find high-quality solutions at scale. To address this, we employ genetic algorithms (GAs)—a class of metaheuristic search methods inspired by natural evolution—to efficiently explore high-dimensional design spaces and balance competing objectives (e.g., maximizing user throughput while maintaining robust sensing performance). This approach is particularly effective in multi-antenna scenarios (MIMO/MISO). 
Unlike gradient-based methods, which can become trapped in local optima despite their faster convergence in non-convex scenarios, GAs leverage population-based data and evolutionary principles—such as crossover and mutation—to explore a broader solution space. This global search capability  increases the likelihood of finding a global optimum, leading to notable enhancements in network performance and overall wireless efficiency \cite{ga_33,gahaps2}.

\subsection{Multi-Objective Optimization with NSGA-II}
In many real-world scenarios, such as HAPS-ISAC, multiple objectives (e.g., improving communication and enhancing sensing accuracy) coexist and often conflict.
To manage this, the Non-dominated sorting GA II (NSGA-II) is employed. NSGA-II is a powerful multi-objective optimization algorithm capable of generating a diverse and well-distributed set of "Pareto-optimal solutions" in a single run. A Pareto-optimal solution is one where no objective can be improved without sacrificing another. By providing a comprehensive set of trade-off solutions, NSGA-II allows engineers and system designers to select the most suitable configuration based on specific mission priorities. This approach is particularly valuable for dynamic environments, as it offers a holistic view of all feasible trade-offs, enabling flexible decision-making.
\section{HAPS-ISAC: A Practical Roadmap Towards 6G Deployment}
\label{futureee}
The successful deployment of HAPS-ISAC networks for 6G hinges on overcoming several practical challenges.
This section provides a roadmap that transitions our theoretical and simulation-based findings into a feasible
implementation plan.
\subsection{Scalable Architectures and Adaptive Resource Management}
The proposed HAPS-ISAC framework supports scalable deployment by allowing integration of
additional HAPS units to accommodate growing traffic and sensing needs. Future research should
explore adaptive resource management strategies that respond to real-time changes in user density,
mobility patterns, and environmental dynamics,
ensuring seamless and efficient network expansion.

\subsection{Robustness to Environmental Dynamics and Channel Imperfections}
Environmental factors such as weather variations, terrain-induced shadowing, and signal blockage remain critical challenges. Future studies should investigate robust optimization techniques and dynamic system reconfiguration methods that maintain service continuity and quality under harsh or unpredictable conditions. This includes developing more accurate and adaptive channel models for diverse
propagation environments.

\subsection{Advanced Machine Learning Integration for Autonomous Operations}
Incorporating machine learning offers strong potential for predictive beam management, real-time
user tracking, and autonomous decision-making. Research should address the design of
lightweight and online-learning models suitable for HAPS-UAV coordination, particularly under
resource-constrained conditions,
enabling intelligent and self-optimizing aerial networks.

\begin{table}[t]
	\centering
	\footnotesize
	\caption{Key Simulation Parameters for the Dual HAPS-ISAC System Models}
	\label{tab:haps_comparison2}
	\renewcommand{\arraystretch}{1.6}
	\setlength{\tabcolsep}{4pt}
	\setlength{\arrayrulewidth}{0.8pt}  
	\begin{tabular}{|>{\centering\arraybackslash}p{1.7cm}|>{\centering\arraybackslash}p{2.0cm}|>{\centering\arraybackslash}p{1.65cm}|>{\centering\arraybackslash}p{1.65cm}|}
		\hline
		\rowcolor{greenheader}
		\textbf{\textcolor{white}{Parameter}} & 
		\textbf{\textcolor{white}{Description}} & 
		\textbf{\textcolor{white}{System A}} & 
		\textbf{\textcolor{white}{System B}} \\
		\hline
		
		\rowcolor{greenlight}
		HAPS Role & Primary function & CPU & SMBS \\
		\hline		
		
		HAPS Alt. & Altitude & 20 km & 20--50 km \\
		\hline
		
		\rowcolor{greenlight}
		UAV Alt. & UAV altitude & 40 m & N/A \\
		\hline
		
		Users (K) & Communication users & 4/cluster & 4 CUs \\
		\hline
		
		\rowcolor{greenlight}
		Targets (J) & Sensing targets & 4/cluster & 4 targets \\
		\hline
		
		HAPS Ant. & HAPS antenna & 20×20 UPA & 8×8 UPA \\
		\hline
		
		\rowcolor{greenlight}
		UAV Ant. & UAV antenna & 4×4 UPA & N/A \\
		\hline
		
		Access Freq. & User frequency & Sub-6 GHz & S-band \\
		\hline
		
		\rowcolor{greenlight}
		Backhaul & UAV-HAPS freq. & Sub-THz & N/A \\
		\hline
		
		TX Power & Transmit power & Optimized & 52 dBm \\
		\hline
		
		\rowcolor{greenlight}
		Channel & Fading model & N/A & Rician K=10 \\
		\hline
		
		Algorithm & Optimization & NSGA-II & GA \\
		\hline
		
		\rowcolor{greenlight}
		Parameters & Config. & Pop:1700 Gen:5000 & Pop:2500 Gen:1500 \\
		\hline
		
	\end{tabular}
\end{table}
\subsection{Security, Privacy, and Resilience in Integrated Infrastructures}
As HAPS-ISAC systems become tightly integrated into critical infrastructure, securing data and system operations becomes essential. Future work must explore  
AI-enabled threat detection, end-to-end encryption schemes, and resilience against cyber-physical attacks tailored to aerial platforms,
ensuring the integrity and trustworthiness of 6G services.

\subsection{Deployment Strategies and Economic Viability}
Scalable adoption of HAPS-ISAC will require economically viable deployment models. Research directions include cost-performance tradeoff analysis, multi-stakeholder coordination frameworks, and practical strategies for infrastructure sharing across communication and sensing operators, facilitating widespread and sustainable implementation.

To overcome the high latency of traditional optimization methods like metaheuristic algorithms, the proposed roadmap involves training  AI models on the optimal solutions generated by these powerful techniques, enabling real-time resource allocation with significantly reduced computational cost.

The roadmap proposes the integration of reconfigurable intelligent surfaces (RIS) to mitigate signal blockages in dense urban areas, leading to a significant improvement in signal quality in alignment with 3GPP NTN standards.

Finally, the roadmap prioritizes developing robust algorithms to counteract channel and sensing uncertainties, thereby ensuring stable system performance in realistic operating conditions.

\section{CASE STUDY: ILLUSTRATING HAPS' DUAL ROLE THROUGH PREVIOUSLY
	DEVELOPED FRAMEWORKS}
\label{casestudy}
\begin{figure}[t]
	\centering
	\includegraphics[width=0.48\textwidth]{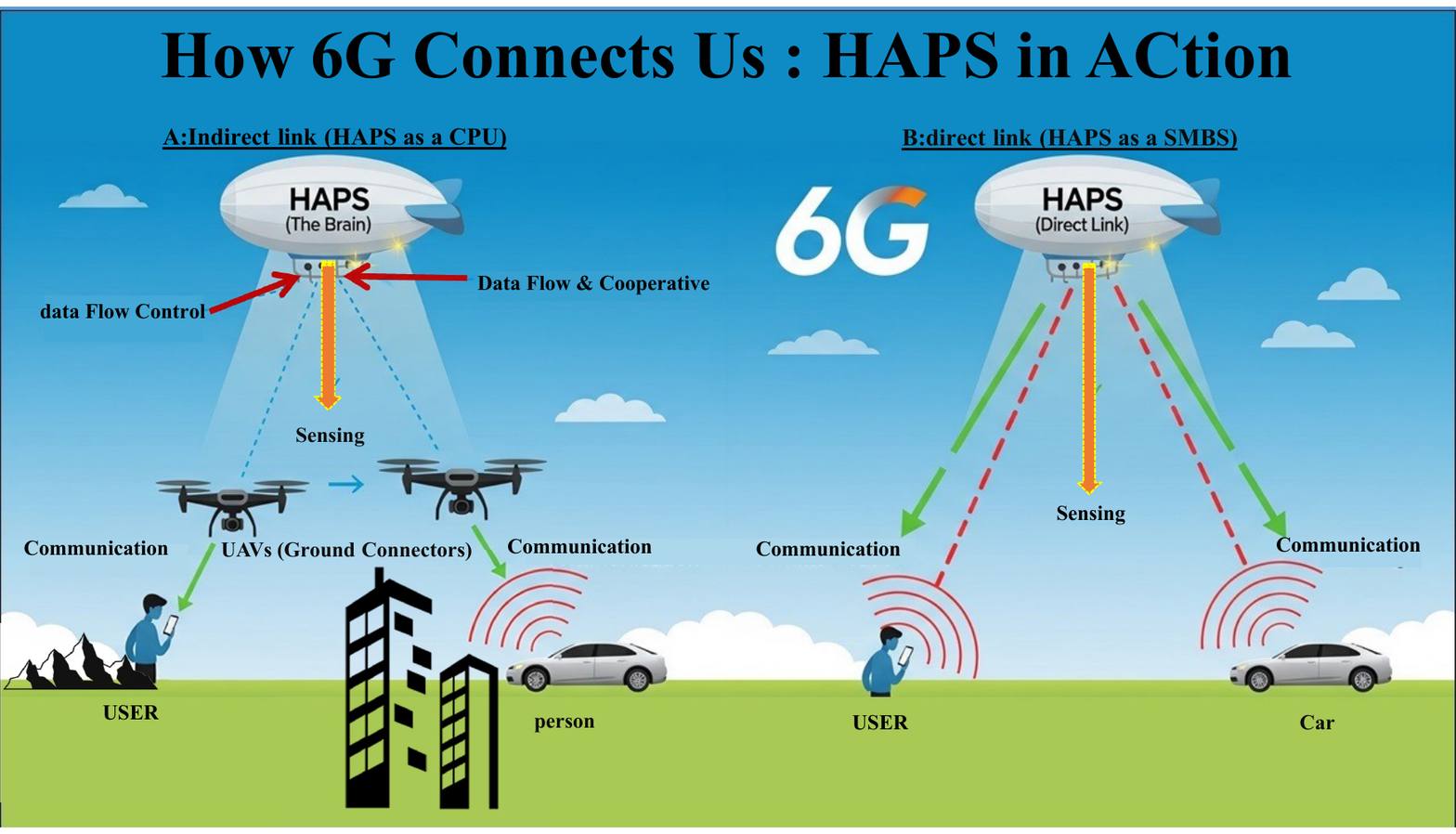}
	\caption{Architecture of HAPS-ISAC systems with two models: (A) HAPS-UAV cooperative ISAC, and (B) HAPS-based monostatic ISAC, both supporting downlink communication and sensing capabilities for multiple users and various ground targets.}
	
	\label{systemm}
\end{figure}

In this section, we evaluate the performance and effectiveness of HAPS-ISAC systems by
presenting two distinct case studies, each illustrating a key operational mode of HAPS: its role as
a CPU within a HAPS-UAV cooperative ISAC system, and its function
as a standalone macro base station in a direct HAPS-ISAC scenario.
These case studies draw upon foundational frameworks and simulation results previously
developed by our research group \cite{ojcom_kh,TWC_kh}, adapted and contextualized here to demonstrate the
versatility and comprehensive capabilities of HAPS-ISAC as envisioned for 6G networks. As
shown in Fig.~\ref{systemm} , HAPS can interact either indirectly via UAVs (System A) or directly with ground
 CUs and sensing targets (System B).

The Table \ref{tab:haps_comparison2} summarizes the core parameters for the
two case studies: System A, where HAPS acts as a CPU for UAV-enabled ISAC, and System B, where HAPS operates as a standalone
SMBS. The parameters are based on the simulation scenarios detailed in the paper.
 \subsection{HAPS-UAV enabled ISAC (HAPS as CPU)}
\label{88}
System~A implements a hierarchical and cooperative architecture in which a high-altitude HAPS functions as a central controller, while multiple low-altitude UAVs operate as agile edge nodes. As illustrated in Fig.~\ref{systemm}, the UAVs are responsible for direct communication with nearby users and localized sensing of ground targets, whereas the HAPS manages centralized coordination, data aggregation, and computational processing.

A key enabler of this architecture is its dual-band operation across two communication domains: the UAV-to-ground access links operate in the Sub-6\,GHz band to ensure reliable coverage, while the UAV-to-HAPS backhaul links utilize the Sub-THz band (e.g., 120\,GHz), offering high-capacity and low-latency transmission of sensing data for centralized processing.

The core challenge in this setup is to balance the inherently conflicting goals of communication quality and sensing accuracy. 
This challenge is addressed via a multi-objective optimization problem aiming to simultaneously maximize (i) the minimum SINR across all  CUs, and (ii) the received sensing signal power at the HAPS, which serves as a proxy for detection capability.
Enhancing one objective often compromises the other due to total power constraints. For instance, given the limited transmit power budget, allocating more power to distant sensing targets inevitably reduces the power available for nearby  CUs, thereby degrading their SINR. Conversely, prioritizing high minimum SINR for all CUs by increasing their allocated power reduces the sensing power, which impairs target detection precision.
To solve this complex, non-convex problem, we employ the NSGA-II. Unlike conventional single-objective methods, NSGA-II is designed to explore the entire Pareto front in a single execution. This provides a diverse set of non-dominated solutions, enabling application-specific choices based on system-level trade-offs. For a mission prioritizing wide-area surveillance, a solution with higher sensing gain but lower average SINR might be selected. Conversely, an application like autonomous vehicle coordination requiring 
ultra-reliable low-latency communication
 (URLLC) would necessitate a solution from a different Pareto point that favors higher SINR. This methodology transforms the optimization challenge from finding a single "best" solution to identifying a comprehensive set of viable operational trade-offs.

The simulation scenario features a HAPS operating at 20\,km altitude and multiple UAVs flying at 40\,m.
 Each UAV serves a cluster comprising four ground CUs and four ground sensing targets. The backhaul link is configured at 120\,GHz, with directional beamforming supported by a \(20 \times 20\) UPA at the HAPS and \(4 \times 4\) UPAs on the UAVs. Access links to users operate in the Sub-6\,GHz band. 
To evaluate the performance of the system under the defined objectives, NSGA-II is configured with a population size of 1700 and 5000 generations.
 This extensive simulation framework enables the exploration of complex trade-off surfaces and reveals valuable insights into the interaction between communication efficiency and sensing effectiveness under the cooperative HAPS-UAV ISAC paradigm.

 \begin{figure}[t]
	\centering
	\includegraphics[width=0.4\textwidth]{ 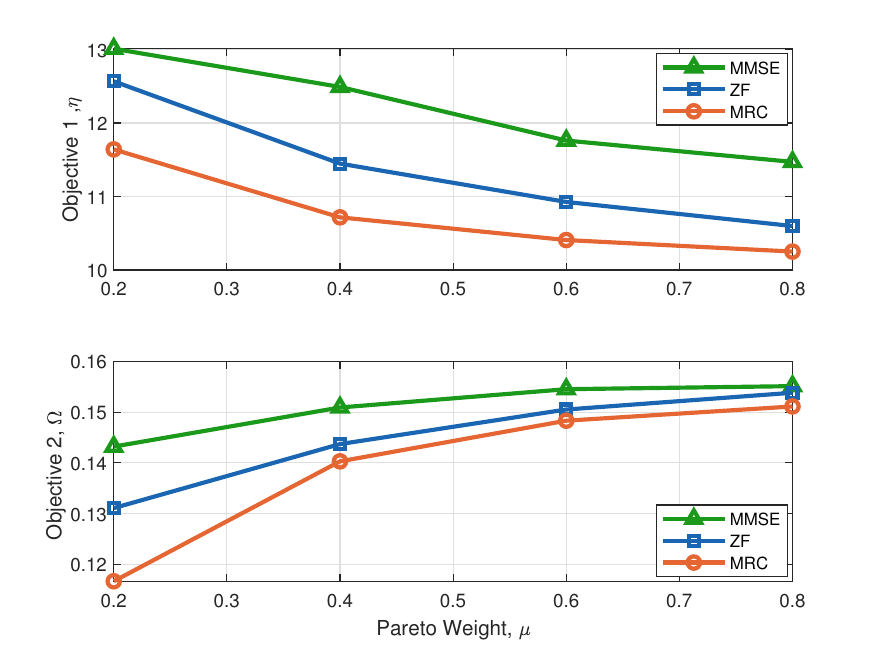}
	\caption{
		Two-dimensional Pareto-optimal front illustrating the trade-off between the two objective functions as  $\mu$ varies for MMSE, ZF and MRC decoders at the CUs.}
	\label{fig1_1}
\end{figure}

Fig. \ref{fig1_1} shows the Pareto curves for CUs with multiple antennas using zero-forcing (ZF), minimum mean square error (MMSE), and maximum ratio combining (MRC) decoders. The trade-off between objectives $\eta$ (worst-user SINR) and  $\Omega$ (sensing quality) is clearly seen, where increasing $\mu$ leads to a decrease in $\eta$ and an increase in $\Omega$, illustrating the inherent conflict between the objectives.
Additionally, MMSE outperforms ZF, which in turn outperforms MRC, due to MMSE’s superior ability to reduce noise and interference. ZF excels in interference suppression, while MRC performs better in high-noise scenarios due to its robustness.

%

 \begin{figure}[t]
	\centering
	\includegraphics[width=0.42\textwidth]{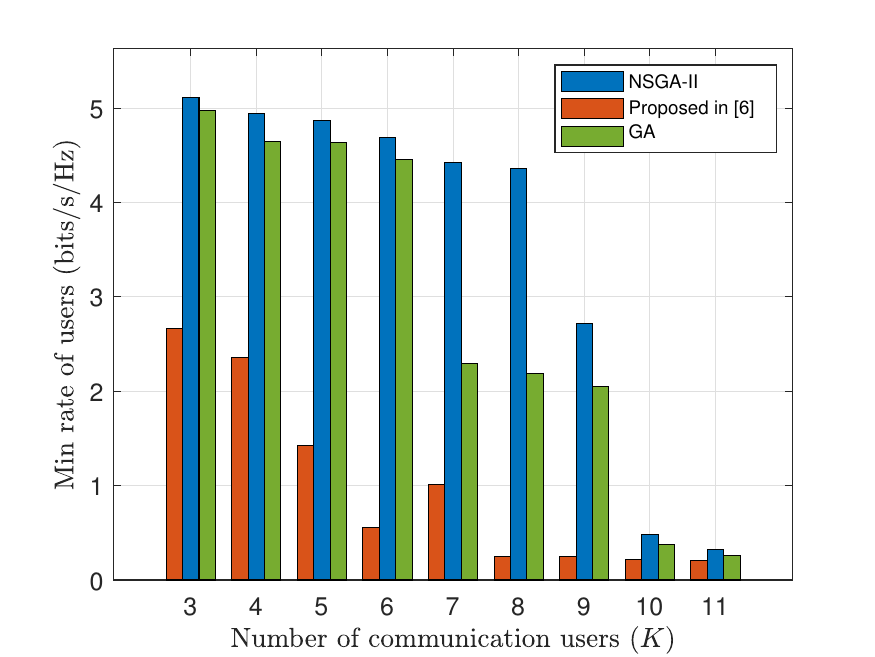}
	\caption{Comparison of minimum user rates between the proposed model and that in \cite{man} for varying user numbers $K$ in a HAPS-UAV-enabled ISAC system.}
	\label{fig2_2}
\end{figure}
A comparative analysis is conducted against the UAV-based system in~\cite{man}, which employs a system model similar to ours. As depicted in Figure~\ref{fig2_2}, our proposed HAPS-based framework yields a higher minimum user rate, demonstrating superior fairness compared to the weighted sum-rate maximization approach of~\cite{man}. For a fair benchmark, the optimization problem in~\cite{man} was solved using the same GA applied in our work. Unlike the communication-centric objective in~\cite{man}, our model prioritizes fairness by maximizing the minimum beampattern gain for sensing targets while maintaining a predefined quality of service (QoS) for CUs. These findings confirm that our integrated HAPS, UAV, and ISAC architecture significantly improves both user fairness and system performance, highlighting its efficacy for enhancing the efficiency of future wireless networks.


	\subsection{HAPS-ISAC System (HAPS as Super  Macro Base Station))}
	System B models the HAPS as a high-altitude SMBS executing monostatic ISAC operations. As illustrated in Fig.~\ref{systemm}, the HAPS simultaneously delivers downlink communication services to multiple ground CUs while performing radar-based sensing of potential targets. This dual capability is enabled by a UPA, whose beams are dynamically steered to optimize both communication quality and sensing accuracy.
	At the heart of this system lies a fairness-driven optimization framework, designed to maximize the weakest sensing quality---quantified by the minimum beampattern gain toward the set of predefined target directions. This objective is subject to two essential constraints: (i) satisfying a minimum SINR threshold for all CUs, and  (ii) adhering to a total transmit power limit \(P_{\max}\). The underlying wireless channels are modeled using a Rician fading distribution, effectively capturing both LoS and non-LoS (NLoS) characteristics inherent to high-altitude platforms.
	Due to the tightly coupled design of the beamforming vectors and the non-convexity of the optimization landscape, the problem is inherently NP-hard and not amenable to conventional convex solvers. To address this challenge, a GA is adopted---a well-established metaheuristic approach known for its scalability and global search capability in high-dimensional spaces.
	 The GA is implemented with a population size of 2500, 1500 generations, a high crossover fraction, and Gaussian mutation with controlled variance. These parameters are carefully tuned to ensure a balance between exploration and convergence, ultimately producing robust and high-quality solutions to the joint communication-sensing beamforming problem.
	To evaluate system performance, a detailed simulation scenario is configured. A single HAPS equipped with an \(8 \times 8\) UPA operates at 20 km altitude in the S-band (2.545 GHz), serving four single-antenna CUs and tracking four distinct ground targets. The total transmission power is capped at 52 dBm, and the channel model assumes a Rician distribution with a high K-factor (=10) to reflect dominant LoS conditions.

\begin{figure}[t]
	\centering
	\includegraphics[width=0.4\textwidth]{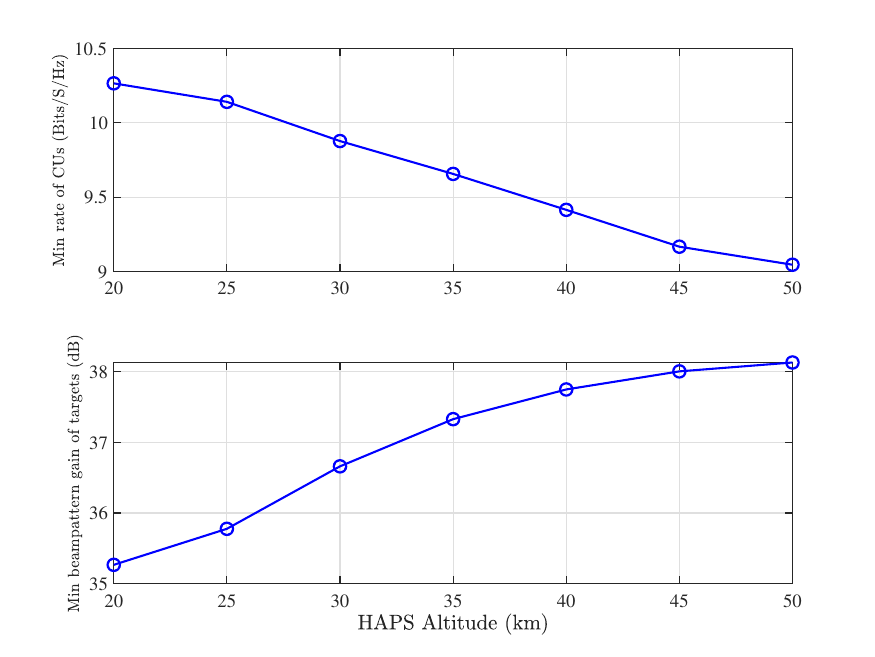}
	\caption{Minimum beampattern  gain and Minimum SINR versus HAPS altitude.}
	\label{fiatude_beam}
\end{figure}
Figure \ref{fiatude_beam} illustrates a fundamental performance trade-off in HAPS-ISAC systems, governed by the platform's operational altitude.
 As altitude increases, the minimum beampattern gain consistently improves. While improved LoS conditions may also contribute, this improvement is primarily due to the geometric clustering of users and targets within a narrower angular spread as seen from the HAPS. Under a fixed total-power constraint, this concentration allows the beamforming optimizer to allocate energy more efficiently, creating a sharper main lobe that simultaneously covers multiple directions.
 In contrast, the minimum SINR for CUs shows a monotonic decline with altitude. This degradation is attributed to two main factors: (i) increased free-space path loss due to the greater HAPS-user distance, and (ii) elevated spatial correlation among user channels, a direct consequence of the angular clustering. This spatial correlation significantly reduces the effectiveness of interference-mitigation techniques such as zero-forcing precoding, thereby impairing communication link quality.
 
 This trade-off establishes altitude as a critical, application-driven design parameter. Lower altitudes are optimal for communication-centric scenarios (e.g., broadband backhaul), whereas higher elevations are superior for sensing-centric missions (e.g., wide-area surveillance). Consequently, these findings motivate treating altitude not as a fixed attribute but as a tunable parameter. Future work should focus on developing dynamic altitude-control loops integrated within joint beamforming and resource-allocation frameworks. Such a system would enable the HAPS to adapt its position in real-time, continuously steering toward a Pareto-optimal altitude that best satisfies instantaneous service demands.

%
%

\begin{figure}[t]
	\centering
	\includegraphics[width=0.36\textwidth]{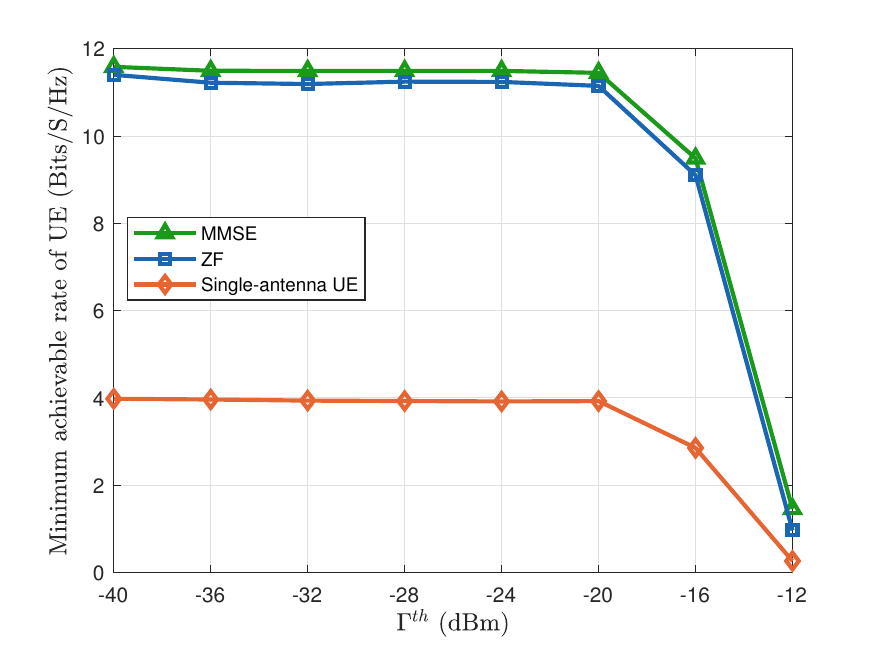}
	\caption{Minimum achievable rate for CUs versus the beam pattern gain threshold.}
	\label{gama_fig}
\end{figure}

With the evolution toward 6G networks, the demand for ultra-high data rates, enhanced spectral efficiency, and ultra-reliable low-latency communication has intensified. To address these stringent requirements, modern user equipment (UE)—such as advanced smartphones—is increasingly equipped with multiple antennas, enabling capabilities such as beamforming, spatial multiplexing, and massive MIMO. Building on this trend, our HAPS-UAV ISAC framework is extended to incorporate multi-antenna CUs, aligning with the performance expectations of future networks.
Accordingly, in our simulations, CUs are modeled as multi-antenna devices. To evaluate receiver performance, two common decoding strategies are considered: ZF and MMSE. These configurations are compared against a baseline single-antenna scenario to quantify the performance benefits of spatial diversity.
Figure~\ref{gama_fig} presents the effect of increasing beam pattern gain thresholds on the worst-case SINR experienced by different user configurations. Three scenarios are compared: single-antenna users, and multi-antenna users employing either MMSE or ZF decoding. The results show a clear trade-off between enhanced sensing constraints and communication quality—tightening the beamforming gain requirement leads to a noticeable drop in the minimum SINR, particularly for users in challenging channel conditions. Nevertheless, multi-antenna configurations consistently outperform their single-antenna counterparts, benefiting from improved spatial diversity and more effective interference suppression.

\section{Discussion}

This section presents the core innovations of the proposed HAPS–ISAC framework for 6G, highlights its advantages over existing approaches, and discusses practical deployment considerations.

\subsection{Innovation and Architectural Advantages}

%
A core innovation of this work is a hierarchical (or tiered) HAPS-ISAC architecture. In this structure, the HAPS not only functions as a communication platform but also acts as a CPU—a "central brain"— for auxiliary UAVs. This design allows computationally intensive tasks to be offloaded from energy-constrained UAVs to the HAPS, significantly extending their operational endurance (i.e., their active mission time). This capability is crucial for applications like persistent surveillance or emergency response.
Furthermore, this study introduces a resource optimization strategy centered on Max-Min Fairness. Unlike conventional approaches that solely maximize the total system data rate, our method guarantees a minimum QoS for all CUs and sensing targets, even those in poor channel conditions. This focus on fairness is vital for achieving the reliable and inclusive connectivity envisioned for 6G.

\subsection{Comparative Analysis and Algorithmic Effectiveness}

The proposed HAPS–ISAC framework outperforms conventional UAV-only ISAC systems in two key areas: user throughput and fairness of resource allocation, as demonstrated by our comparative evaluations. This advantage stems from HAPS’s broader coverage, higher transmit power, and its centralized processing role.
From an optimization perspective, NSGA-II is well suited to the multi-objective, non-convex nature of the problem. Unlike reinforcement learning methods such as Proximal Policy Optimization (PPO), which typically require scalarization and may overlook portions of the Pareto front, NSGA-II directly explores a diverse set of Pareto-optimal solutions. This flexibility enables decisions tailored to specific network priorities and significantly improves adaptability under dynamic conditions.
\subsection{Practical Implementation Challenges and Future Directions}
While our framework shows strong theoretical performance, its practical deployment faces challenges in computational complexity, channel modeling, and network security.

First, the system's scale makes optimization computationally demanding. Although GAs offer a robust heuristic, their real-time performance can be enhanced by techniques such as \textbf{warm-starting} with quality initial solutions, \textbf{parallelization} of fitness evaluations, and creating \textbf{hybrid algorithms} that combine GAs with faster local search methods.

Second, existing channel models must be extended beyond simple LoS assumptions. Future models need to incorporate the effects of \textbf{NLoS paths}, prevalent in urban areas, and manage the \textbf{jitter and mobility} introduced by the quasi-stationary HAPS and mobile ground users through dynamic control loops.

Finally, the integrated system presents a new attack surface, making security paramount. Future work must focus on developing lightweight, \textbf{AI-driven threat detection} models, especially for constrained devices like UAVs, and implementing robust \textbf{end-to-end encryption} to secure both communication and sensing data.

 \section{CONCLUSION}
\label{concl}
This article introduces a novel and efficient HAPS-ISAC framework tailored to meet the stringent
dual communication and sensing demands of 6G networks. By leveraging
HAPS's unique advantages—including wide-area coverage, high energy efficiency, and reliable
LoS connectivity—the proposed architecture enables simultaneous optimization
of communication and sensing performance in complex environments. Simulation results
demonstrate significant improvements in key performance metrics, including user throughput,
SINR, sensing accuracy, and fairness in resource allocation, particularly when compared with purely
UAV-based systems. 
A key feature of the proposed framework is the dual-role deployment of HAPS: it serves as an SMBS providing wide-area, high-power coverage and as a CPU for auxiliary UAVs. Offloading compute-intensive tasks from energy-constrained UAVs to the HAPS extends flight endurance and reduces end-to-end latency.
These findings validate the strong potential of HAPS-ISAC architectures to
support highly scalable, adaptive, and intelligent aerial networks for 6G. Future work will focus
on advancing real-time system reconfiguration using sophisticated machine learning techniques,
integrating robust security protocols, and developing highly accurate models for complex aerial-terrestrial
channels to enable practical and widespread deployments in next-generation wireless
systems.

\bibliographystyle{ieeetr}
\bibliography{ref_competiotion.bib}

\end{document}